%% file: main.tex
\documentclass[manuscript]{acmart}
\usepackage{tabularx}
\usepackage{hyperref}
\usepackage{graphicx} 
\usepackage{subcaption}

\AtBeginDocument{%
  \providecommand\BibTeX{{%
    \normalfont B\kern-0.5em{\scshape i\kern-0.25em b}\kern-0.8em\TeX}}}

\newcommand{\italquote}[1]{\begin{quote}``\textit{#1}''\end{quote}}
\newcommand{\redacted}[1]{\textbf{REDACTED}} 

\setcopyright{acmcopyright}
\copyrightyear{2018}
\acmYear{2018}
\acmDOI{10.1145/1122445.1122456}

\begin{document}

\title{Understanding User Perspectives on Prompts for Brief Reflection on Troubling Emotions}
\author{Ananya Bhattacharjee}
\email{ananya@cs.toronto.edu}
\affiliation{%
  \institution{Computer Science, University of Toronto, Canada}
}


\author{Pan Chen}
\email{pan.chen@utoronto.ca}
\affiliation{%
  \institution{Computer Science, University of Toronto, Canada}
}

\author{Linjia Zhou}
\email{linjia.zhou@mail.utoronto.ca}
\affiliation{%
  \institution{Cognitive Science, University of Toronto, Canada}
}

\author{Abhijoy Mandal}
\email{a.mandal@mail.utoronto.ca}
\affiliation{%
  \institution{Computer Science, University of Toronto, Canada}
}

\author{Jai Aggarwal}
\affiliation{%
  \institution{Computer Science, University of Toronto, Canada}
}

\author{Katie O'Leary}
\affiliation{%
  \institution{ Google, Inc, Seattle, Washington, United States}
}

\author{Anne Hsu}
\affiliation{%
  \institution{Queen Mary University, London, United Kingdom}
}

\author{Alex Mariakakis}
\affiliation{%
  \institution{Computer Science, University of Toronto, Canada}
}
\author{Joseph Jay Williams}
\affiliation{%
  \institution{Computer Science, University of Toronto, Canada}
}



\begin{abstract}
\input{chapters/0-abstract}
\end{abstract}



\maketitle

\input{chapters/1-intro}
\input{chapters/2-related}
\input{chapters/3-rqa}
\input{chapters/4-mturk-study}
\input{chapters/5-comparison-study}
\input{chapters/6-longitudinal-study}
\input{chapters/7-discussion}
\input{chapters/8-conclusion}


\bibliographystyle{ACM-Reference-Format}
\bibliography{sample-base}
\end{document}

%% file: chapters/0-abstract.tex
We investigate users’ perspectives on an online reflective question activity (RQA) that prompts people to externalize their underlying emotions on a troubling situation. Inspired by principles of cognitive behavioral therapy, our 15-minute activity encourages self-reflection without a human or automated conversational partner. A deployment of our RQA on Amazon Mechanical Turk suggests that people perceive several benefits from our RQA, including structured awareness of their thoughts and problem-solving around managing their emotions. Quantitative evidence from a randomized experiment suggests people find that our RQA makes them feel less worried by their selected situation and worth the minimal time investment. A further two-week technology probe deployment with 11 participants indicates that people see benefits to doing this activity repeatedly, although the activity may get monotonous over time. In summary, this work demonstrates the promise of online reflection activities that carefully leverage principles of psychology in their design.

%% file: chapters/1-intro.tex
\section{Introduction}
Most people would welcome help to better manage their stress and negative emotions, and computer-mediated communication (CMC) platforms make it easier than ever before for people to access resources to address this desire.
Beyond social media groups and text messaging programs that are designed to help specific populations deal with challenges relevant to their context (e.g., college life~\cite{amanvermez2021stress}, occupation-specific issues \cite{george2013facebook}), online programs like \textit{Beating the Blues}~\cite{proudfoot2003computerized} have been proposed to help people learn and apply principles of psychological theory to help them deal with their emotions.
Although research has shown that participating in online programs can yield similar benefits to what can be garnered from in-person therapy~\cite{cuijpers2008internet, proudfoot2003computerized}, engaging in them requires committing to a series of hour-long sessions to receive their full effect.
Convenience is an important factor in people's willingness to use online resources~\cite{bernecker2017web}, so the significant time investment required for online mental health programs has led many people to quit them early~\cite{eysenbach2005law}. 

In this work, we begin to explore a brief online activity that people can complete to help them manage their negative emotions.
The guiding principles of our exploration are as follows:
\begin{itemize}
    \item \textbf{Minimal time commitment:} The activity should be simple enough so that people can complete it in 15 minutes --- the equivalent of a midday coffee break at work or a fraction of a person's morning routine.
    \item \textbf{Applicability to the general public:} The activity should not be targeted towards a particular domain, culture, or population. In other words, the activity should be generalizable to the point where people can adapt it to their own context and situation.
    \item \textbf{Scalability:} The activity should be implemented and deployed in a way that does not need a significant support structure. This means that the activity should not require a live conversational partner or intensive scaffolding (e.g., tutorial videos).
\end{itemize}
 
Previous work in education suggests that posing carefully crafted questions could be useful for prompting deeper reflection and understanding, even without live feedback from a conversational partner~\cite{williams2016revising, slovak2017reflective}.
With that in mind, we designed a digital reflection question activity (RQA) that asks people to reflect on a troubling situation. 
The questions in our RQA help people articulate their thoughts and emotions about the situation using principles from cognitive behavioral therapy (CBT)~\cite{beck2011cognitive}.
We were particularly inspired by thought records~\cite{greenberger1995mind} and behavioral chaining analysis~\cite{rizvi2014mastering} --- techniques that encourage people to connect their thoughts, experiences, and emotions to identify triggers that generate negative patterns and come up with alternative ways of thinking.

We provide insights into the design of our RQA and how it was experienced by users, which we hope will inform the design of future interventions with similar goals. 
We gathered these observations through three studies.
In our first study, we administered the RQA through Amazon Mechanical Turk (MTurk) to gather qualitative feedback on the activity's design.
In our second study, we again administered the RQA through MTurk, but this time to measure the instantaneous impact that the RQA could have relative to a baseline activity in which people were asked to write about a troubling situation in response to a single question.
In our third and final study, we observed the longitudinal impact that the RQA could have by repeatedly delivering it to people over the course of two weeks; we also explored the implications of distributing the RQA over email versus text messaging.

We found that the structured analysis supported by our RQA helped people reduce their stress and identify solutions for improvement. 
Although our RQA consisted of nine questions, people did not complain about the time commitment required to complete it and generally wrote thoughtful responses to the prompts.
However, deploying the RQA over the course of two weeks raised some potential challenges, including the monotony of doing the same activity several times, the limited affordances of mobile phones, and the importance of having the prompts align with the occurrence of new troubling situations. 
These highlight design considerations and opportunities for the broad range of RQAs HCI practitioners and researchers can deliver through a range of CMC platforms, such as giving users control over the frequency of prompts and automated question personalization.

In summary, our main contributions are:
\begin{itemize}
    \item The creation of a prototype brief digital reflective question activity (RQA) that people can complete on their computer or phone to reflect upon a stressful situation,
    \item Insights gathered via surveys and interviews from 122 people who engaged with the RQA through MTurk,
    \item Confirmation of the benefits of the RQA relative to a baseline activity via an comparison study run on MTurk with 215 participants, 
    \item Observations and design considerations from a two-week deployment of our RQA using different platforms.
\end{itemize}

%% file: chapters/2-related.tex
\section{Related Work}
In this section, we discuss how researchers in HCI and psychology alike have developed interventions to support mental health and psychological well-being.
Programs and interventions in this domain are often centered around a psychological framework.
We focus our attention on cognitive behavioral therapy (CBT) --- a family of clinical psychology approaches that help people to identify the relationships between their thoughts and distressing emotions~\cite{beck2005current}. 
CBT has been shown to be at least as effective as medication in treating mental health issues ranging like depression, anxiety, and addiction \cite{butler2006empirical}, making it a staple of clinical psychology programs. 

We organize our prior work by first describing activities that require a significant time commitment and then activities that can be done in a shorter amount of time.
Lastly, we provide an overview of how researchers have developed interventions centered around principles of self-reflection.

\subsection{Supporting Mental Health via Extended Engagement}
When people think of mental health support, they most often recall face-to-face interactions between professional counselors and their clients. 
These interactions are beneficial and irreplaceable in many situations, but they can require significant time investment from both parties~\cite{buyruk2019early}. 
The stigma around seeking mental health support and fear of face-to-face interactions are among other reasons that prevent people from seeking support from a trained professional~\cite{dinapoli2014theory}.

The ubiquity of computers and mobile phones enables people to access resources anytime and anywhere, so researchers have explored how CMC platforms can be used to deliver mental health support.
There has been a proliferation of online applications that deliver CBT-related interventions~\cite{deady2016online,anttila2020web,cuijpers2008internet,cuijpers2008internet,gregory2004cognitive}, both for people with clinically diagnosed mental health disorders and for the general population.
Many of these interventions take the form of websites and apps.
For example, \citet{deady2016online} developed a 4-week web-based program intended to help participants reduce their alcohol consumption; the program was divided into a series of one-hour modules that were meant to be completed one week at a time. 
Similarly, \citet{anttila2020web} created a series of 45-minute web modules targeted towards adolescents with symptoms of depression.
Other online programs, such as Beating the Blues \cite{attwood2016exploring}, E-couch \cite{calear2016cluster}, and MoodGym \cite{topper2017prevention}, have also been shown in randomized trials to effectively enhance mental health outcomes and reduce clinical symptoms~\cite{cuijpers2008internet, coull2011clinical}.

The ubiquity of mobile phones has given rise to several text messaging systems as well \cite{aguilera2017automated, fitzpatrick2017delivering, zwickert2016high}.
\citet{aguilera2017automated} created a text messaging system to supplement a weekly face-to-face group meeting among low-income Latinos with depressive symptoms.
The system included messages designed to be sent at a variety of intervals: daily (mood self-assessments, supplements to live therapy content, medication reminders), weekly (meeting reminders), and monthly (opportunities to opt-out).
To investigate a standalone text messaging intervention, \citet{fitzpatrick2017delivering} developed a two-week intervention program in which college students with depressive symptoms interacted with a conversational agent named Woebot. 
The chatbot prompted participants to engage in a new conversation each day, following a list of questions inspired by CBT principles. 

Despite the major benefits to be gained by the aforementioned online interventions, many of them still require substantial time commitment from users, whether it be in the form of lessons, videos, or homework activities. 
These programs are also designed to last multiple weeks in order to cover all of the content they have to offer.
Like free online educational courses~\cite{yang2013turn}, many people are eager to start these programs but later drop out due to busy schedules or a lack of energy~\cite{eyal_2019}.
Therefore, there is a growing call in the HCI community to design alternative online interventions that do not require significant effort from users across an extended period of time~\cite{schueller2013realizing}.


\subsection{Supporting Mental Health via Brief Interventions}
Brief activities that require minimal effort and provide tangible benefits can increase self-efficacy and serve as a stepping stone to more extensive treatments~\cite{coyle2009evaluation}.
This opportunity has been recognized by researchers in both clinical psychology and HCI.
From literature in clinical psychology, \citet{murgraff2007reducing} demonstrated that a small two-page pamphlet could help curb unhealthy drinking behaviors amongst female university students, and \citet{carney2020adaptation} used a similar intervention to support substance-using adolescents and their caregivers. 
Another form of brief interventions uses imaginal exposure, wherein a person is shown an image that elicits a memory; this process is intended to encourage people to share their experiences and relieve stress~\cite{bryant2003imaginal}.
\citet{steenkamp2011brief} used imaginal exposure of upsetting experiences to help military service members deal with posttraumatic stress disorder. 
Other examples of brief interventions in clinical psychology can be found in work by \citet{beacham2020brief} and \citet{lyon2014case}, among others.
From literature in human-computer interaction, \citet{elefant2017microinterventions} created a website that gave people access to micro-interventions like breathing exercises and mood self-assessments.
In a randomized controlled trial, they found that engaging with the website led to immediate improvements in people's mood.
PopTherapy by \citet{paredes2014poptherapy} also proposed giving people access to a multiple micro-interventions for stress reduction; however, they used smartphone sensor features (e.g., location, motion) and a machine learning model to recommend specific activities in a personalized manner.

Most of the aforementioned interventions are designed to be brief, regardless of whether they are conducted verbally, on paper, or on a digital device.
In this work, we supplement the literature by translating a complex exercise from CBT that typically requires introduction from a trained professional into an activity that can be completed without scaffolding in a short amount of time.
Unlike the aforementioned work, our activity is design to be flexible enough so that people can reflect on both a recent event from the previous day or an event from the distant past that has not yet been resolved.


\subsection{Self-Reflection}
One particularly important concept in CBT and psychology in general is self-reflection --- an individual's conscious attempt to understand and re-evaluate their own thoughts about a troubling situation, typically in a structured manner~\cite{sedig2001role, malacria2013skillometers}.
Self-reflection is often the driving force that converts one's intentions to action~\cite{baumer2015reflective}.
Moreover, self-reflection allows an individual to view a situation from a different perspective, thus enabling them to understand others' opinions~\cite{diener1979self}. 

HCI researchers in recent years have incorporated reflective activities into mental health interventions, particularly through mobile phone applications that show users summaries of their mood or physical activity~\cite{bhattacharjee_haque_hady_alam_rabbi_kabir_ahmed_2021, lin2006fish, consolvo2008flowers}. 
Self-monitoring tools like these have been found to help people identify the root cause of their issues and motivate them to take necessary actions, even without any explicit recommendations~\cite{bhattacharjee_haque_hady_alam_rabbi_kabir_ahmed_2021}.
Other digital tools have attempted to promote self-reflection through conversational agents~\cite{baumer2014reviewing}.
For example, Wysa~\cite{inkster2018empathy} asks questions like ``What has been the major event or change in your life recently?'' or ``Is it getting hard for you to cope with your daily tasks?'' in order to prompt people to reflect on their struggles.
Another chatbot by \citet{kocielnik2018reflection} encouraged people to engage in healthy behavior by sending them graphs and charts of their daily activity level. 
Similar conversational agents have been deployed to promote healthy drinking habits~\cite{mason2014text} and psychoeducation~\cite{tielman2017should}.
Although chatbots continue to become more sophisticated in parallel with advances in natural language processing, they are still limited in their ability to have nuanced and empathetic conversations.
Furthermore, literature suggests that back-and-forth conversations are not always necessary to elicit self-reflection, as asking probing questions with the words ``why'' or ``how'' can be enough to increase one's own understanding of a problem~\cite{fleck2010reflecting, williams2016revising, bhattacharjee_haque_hady_alam_rabbi_kabir_ahmed_2021}.

Numerous works have attempted to conceptualize a structure for self-reflection~\cite{atkins1993reflection, gibbs1988learning}.
In this work, we attempt to adapt a popular CBT exercise for supporting self-reflection called a thought record~\cite{greenberger1995mind} into an activity that people can complete on their computer or phone.
A thought record is a worksheet with a grid that includes five columns: Situation, Thoughts, Emotions, Behaviors, Alternative Thoughts.
The goal of the exercise is to encourage behavioral chaining~\cite{rizvi2014mastering} --- a process through which people draw connections between their thoughts and emotions to identify triggers and irrational thoughts, revealing potential opportunities to reframe their way of thinking~\cite{atkins1993reflection}. 
Researchers have identified several benefits to thought records and behavioral chaining.
Thought records can help people recall memories of prior events that were initially assumed to be unimportant~\cite{conway2009episodic}. 
Identifying the full timeline of an event can help people recognize their own faulty behavior patterns, thus preparing them for similar events in the future~\cite{williams1986autobiographical}. 
Moreover, informal exposure to negative experiences can increase one's ability to tolerate troubling situations~\cite{bridges2006activating} or recover from problematic behaviors (e.g., binge-drinking, self-harming)~\cite{rizvi2014mastering}. 

Thought records are typically introduced as CBT homework assignments that patients can complete between visits with a trained professional, providing them with the scaffolding to complete the activity.
Our RQA attempts to distill this exercise into a brief guided activity that can be completed on a person's computer or phone without the need for external support.

%% file: chapters/3-rqa.tex
\section{The Design of Our RQA}
\label{sec:design}
We created an RQA inspired by thought records and behavioral chaining to help people reflect on a troubling situation in their lives.
Our research team, which consists of graduate students and faculty members with experience in human-computer interaction and psychology, converged on the design of our activity through an iterative process.
We held multiple sessions to discuss the flow of the activity, the wording of the questions, and the potential reactions that the questions might elicit.
Table~\ref{tab:rqa} shows the nine questions that compose our RQA. 
We first ask the user to think about the details of the situation and to isolate the most important stressor.
Next, we ask the user to consider the thoughts and emotions that are evoked by the stressor.
Since the user will have answered multiple questions at this point, we then ask them to synthesize their responses into a brief summary.
Lastly, we ask the user to reflect on whether their way of thinking is justified or if there is a better way of thinking that they should apply instead.

We presented our RQA to four clinical psychologists with expertise in CBT to validate its construction and to help us consider the best ways to evaluate it.
The psychologists verified that our RQA is aligned with activities that would be used in psychotherapy, but they also remarked that the questions focused on advanced techniques that were usually introduced only after several sessions of evaluation and psychoeducation. 
They also expressed skepticism that people might find the activity too lengthy or that people might not know how to respond to some of the questions. 
One psychologist even posited that more than two questions might be excessive for an online format without a conversational partner. 
These comments motivate the studies that follow in this paper.

\begin{table}[]
    \centering
    \begin{tabular}{|p{0.3\textwidth}|p{0.35\textwidth}|p{0.3\textwidth}|}\hline
        \multicolumn{3}{|p{\textwidth}|}{\textbf{Prompt:} Think of a particular situation where you felt stressed or had a negative emotion, which you can try to reflect on as you go through this activity. It could be a current situation, one in the past, or one you anticipate in the future.} \\ \hline
        \textbf{Question} & \textbf{Example Response} & \textbf{Purpose} \\ \hline
        1. What’s the situation? Feel free to explain it in as much detail as you'd like. & 
        ``My son has moved away and left no way for me to get in contact with him.'' & 
        Provides context for the activity \\ \hline
        
        2. What part of the situation is the most troubling? & 
        ``The fact that he does not care enough to reach out to me and let me know he is safe.'' & 
        Sets an agenda for the rest of the activity \\ \hline
        
        3. What are you thinking to yourself? & 
        ``I hope he is okay and safe. I wonder why he would do this. I thought we had a good relationship.'' & 
        Identifies troubling thoughts \\ \hline
        
        4. What thought is the most troubling? & 
        ``I don't know if he is safe.'' &
        Focuses attention on most the most troubling thought \\ \hline
        
        5. What do you feel when you think this? & 
        ``Panicked and worried.'' &
        Reinforces the core CBT principle that thoughts trigger feelings \\ \hline
        
        6. When you have these feelings, what actions do you take? What actions do you avoid? & 
        ``I try to refocus my thought on something else. I try to avoid thinking about what all bad could be happening to him.'' & 
        Identifies behaviors that are caused by the cascading effect of thoughts and feelings \\ \hline
        
        7. Retype the summary of the situation in the following format:\newline
        Trigger:\newline
        Thought:\newline
        Feeling:\newline
        Behaviour: & 
        ``I am triggered by thoughts of my son taking off and not staying in contact. I think about all the bad things that could happen and why he would do this. I feel panicked and worried. When feeling this way I try to think about other things and not focus on the negative of the situation.'' & 
        Synthesizes past reflection by highlighting the connection between the trigger and its manifestations \\ \hline
        
        8. Consider whether the trigger truly justifies this type of thinking. Explain below. & 
        ``The trigger does justify it. This is my child that I raised. I no longer know where he is, I cannot get in touch with him and I don't know if he is okay.'' & 
        Challenges potentially negative thought patterns \\ \hline
        
        9. If you were to explore an alternative line of thinking, how would you do it? & 
        ``I raised my child to be independent and he is trying to exercise that independence for the first time in his life. He needs me to take a step back for a while so that he can do this on his own.'' & 
        Encourages alternative thoughts that can provoke different feelings and behaviors \\ \hline
    \end{tabular}
    \caption{The prompts and questions that compose our reflective question activity (RQA). The design of these questions is influenced by thought records~\cite{greenberger1995mind} and behavioral chaining~\cite{rizvi2014mastering}.}
    \label{tab:rqa}
\end{table}

%% file: chapters/4-mturk-study.tex
\section{Study 1: User Perspectives After One-Time Use of The RQA}
Our first study was designed to gather feedback on the construction of the RQA irrespective of other factors (e.g., when it was being used, how it compared to other interventions).
Study 1 consisted of two rounds: one for theme elicitation and another for theme validation. 
All participants in this study were recruited prior to the COVID-19 pandemic.

\subsection{Participants}
For the first stage of our study, we recruited 42 participants from MTurk (35 males, 7 females) with an average age of 34.6 years.
We identify these participants as M1--M42.
We also recruited six additional people via email and word-of-mouth from a university campus community (1 male, 5 females) with an average age of 19.7 years.
We identify these participants as L1--L6.
For the second stage of our study, an additional 74 participants were recruited from MTurk (53 males, 21 females) with an average age of 32.3 years.
There were no inclusion criteria since we were interested in observing how our RQA would be perceived by the general population.
Participants were compensated \$4 CAD for completing the RQA and \$15 CAD for being interviewed.

\subsection{Study Procedure}
All participants were asked to complete the RQA online on the Qualtrics survey platform with all the questions on one page.
Participants in the first stage of the study were asked to provide feedback about their experience, which features they appreciated, and which features they thought could be improved.
The crowdworkers were asked to provide this information through a separate online survey with open-ended questions, while the university students participated in semi-structured interviews with similar questions.
The interviews took 45--60 minutes and were held either in-person or through different video-conferencing platforms.
Participants in the second stage of the study were shown the same RQA and then asked to rate their agreement with statements representing the themes that emerged from the previous stage.
The ratings were given along a 7-point scale ranging from ``strongly disagree'' to ``strongly agree''.

\subsection{Data Analysis}
The interview transcriptions and the survey responses from the first stage of the study were analyzed together using a thematic analysis approach~\cite{cooper2012apa}.
After the interviews were transcribed, two researchers examined the data together to familiarize themselves with the general sentiments of the participants. 
The researchers then individually applied the open-coding process~\cite{glaser2016open} to a subset of the data to develop their own preliminary codebooks. 
After sharing their codebooks with one another, the researchers held multiple discussions to consolidate the codes into a shared codebook. 
Next, they applied this codebook to a different subset of the data and again refined the codebook.
Finally, the researchers reached a consensus and applied the final codebook to separate halves of the data. 

\subsection{Ethical Concerns}
\label{sec:study1ethical}
Our research activities were approved by the \redacted{University of Toronto's} Research Ethics Board. 
However, we recognize that conducting research on mental health can raise several ethical issues. 
For example, our particular set of questions can induce stress or symptoms of depression and anxiety, particularly when participants are asked to recall a troubling situation. 
To mitigate this negative outcomes, we clearly explained the potential risks in the consent materials and reminded participants that the RQA was part of a research study. 
We also provided survey participants with the contact information of several mental health helplines.
The interviewers were trained to clearly explain the goal of the project and maintain an appropriate level of empathy and support. 
Interviewers were also trained to run the Columbia-Suicide Risk Assessment protocol~\cite{posner2008columbia} if interviewees showed any indication of self-harm or suicidal ideation.  
Interviewees also had the option to skip any question they did not want to answer or to leave the interview session at any point.

\subsection{Results}
\input{figures/study1-table}
We elicited four major themes during the first stage of our study.
Table~\ref{tab:study1} lists the the themes and shows the degree to which participants in the second stage agreed with them.
We briefly highlight some of the compelling insights from our qualitative data below. 

\subsubsection{Appreciation for Structured Reflection}


Participants were appreciative of the fact that the RQA helped them break down the components of their stressful situation. 
By deconstructing the situation, participants felt that they were able to become more aware of the causes of their negative emotions. 
Some participants also noted that the activity helped them recognize faulty thought patterns. 
For example, M17 said,
\italquote{The activity helped me pinpoint my maladaptive coping strategy ... led me to think more with my brain and less with my immediate emotional reaction.}


\subsubsection{Venting Negative Thoughts Through Writing}

Participants enjoyed expressing their thoughts and feelings through writing as it allowed them to ``get out all thoughts and feelings and take that weight off of my shoulders'' (L5). 
Moreover, some participants appreciated seeing their thoughts typed out in front of them, commenting that the act of writing helped solidify previously nebulous or disjointed thoughts. 
For example, L4 thought that the RQA forced them to dissect their feelings, which otherwise would have been unorganized. 
M11 went on to suggest that writing about their thoughts allowed them to examine their situation ``from an outside perspective''.


\subsubsection{Helping Identify Solutions}
Many participants stated that the activity prompted them to adopt a problem-solving approach to improve their situation. 
They could better identify the root cause of their stress because they were prompted to describe their troubling situation in a structured order, which made it easier for them to find a solution to their problem. 
Since the end of our RQA prompted users to consider alternative ways of thinking, participants like L3 felt empowered since they were often able to emerge from the activity with at least one prototype solution.


\subsubsection{Incidental Negative Side-Effects}
Our RQA received some criticism from the participants who felt even more confused after revisiting their troubling situation.
L5 noted that as they were exploring an alternative line of thinking, they found it confusing to keep track of both their original through process and the reframed one.
Frustration was another negative side-effect that some participants experienced.
L6 felt that the confusion they experienced directly led to frustration, while others were frustrated because they could not identify a solution by the end of the activity. 
As M19 said, ``The questions just made me think about how much pain I was in and really didn't offer any solution whatsoever to the stress.''
Some participants also felt at a loss when asked to think of alternative perspectives on their thoughts.

%% file: figures/study1-table.tex
\begin{table}[]
\centering
\begin{tabular}{|p{0.5\textwidth}|p{0.09\textwidth}|p{0.09\textwidth}|p{0.09\textwidth}|p{0.09\textwidth}|}\hline

\textbf{Theme} & \textbf{Disagree and Neutral} & \textbf{Slightly Agree} & \textbf{Agree} & \textbf{Strongly Agree} \\ \hline
This exercise allowed me to undergo a structured analysis of my situation, which I found helpful & 26\% & 23\% & 38\% & 13\% \\ \hline
It felt good to write out my situation to let it out of my head & 26\% & 22\% & 32\% & 20\% \\ \hline
This activity helped me understand or remember some solutions I could use to improve the situation  & 28\%  &  33\% & 25\% & 14\% \\ \hline
Trying to understand my situation brought more confusion  & 62\%  & 15\% & 12\% & 11\% \\ \hline

\end{tabular}
\caption{The first column shows the themes that emerged from the first round of Study 1, while the remaining columns summarize the degree of agreement that people had with these themes in the second round of Study 1.}
\label{tab:study1}
\end{table}

%% file: chapters/5-comparison-study.tex
\section{Study 2: Comparing The RQA to a Baseline}
When we first showed our RQA to clinical psychologists, one concern that was raised was that giving people more than a couple of questions might be overly burdensome.
With that in mind, our second study was designed to compare the benefits of the RQA relative to a simpler activity wherein participants were asked to write about their stressful situation in response to a single question.
This study allowed us to assess whether the perceived benefits of going through the RQA outweighed the additional time commitment required to answer a series of probing questions.
As with Study 1, all participants in this study were recruited prior to the COVID-19 pandemic.

\subsection{Participants}
We recruited 215 participants from MTurk (133 males, 77 females, and 5 undisclosed) with an average age of 33.8 years. 
As with Study 1, all participants were compensated \$4 CAD for their participation and there were no inclusion criteria. 

\subsection{Study Procedure}
The study had a between-subjects design in which participants were randomized into one of two conditions.
The first condition, which we call \textit{RQA}, entailed participants completing our nine-question reflective question activity.
The second condition, which we call \textit{Baseline}, entailed participants answering the first question of the RQA on its own.
Both conditions asked participants to think through a troubling situation they were experiencing; however, those in the \textit{Baseline} condition were not explicitly asked to reflect on their thoughts and emotions through the additional questions of the RQA. 

\subsection{Analysis}
We collected data before and after participants completed their respective activities to evaluate the following hypotheses:
\begin{itemize}
    \item[\textbf{H1:}] \textbf{(Perceived Benefits)} Participants in the \textit{RQA} condition will experience more instantaneous stress relief from completing the activity than participants in the \textit{Baseline} condition.
\end{itemize}
To evaluate this hypothesis, we asked participants to rate how useful they felt the activity was.
We call this measure \textit{Perceived Utility}, and it was measured along a 7-point scale (-3 for ``strongly disagree'' to 3 for ``strongly agree'').
We also asked participants to rate the degree to which they were feeling troubled about their selected situation before and after the activity.
These ratings were given along an 11-point scale (-5 to +5) to increase the resolution with which people could express their stress.
We call the difference between the ratings before and after activity the \textit{Perceived Stress Change}, with negative values indicating a decrease in stress.
Both \textit{Perceived Utility} and \textit{Perceived Stress Change} were compared across conditions using independent samples t-tests.

\begin{itemize}
    \item[\textbf{H2:}] \textbf{(Elapsed Time)} Participants in the \textit{RQA} condition will take more time to complete the activity than participants in the \textit{Baseline} condition, yet the perceived time commitment will not be significantly different.
\end{itemize}
To evaluate this hypothesis, we recorded the time it took for participants to complete the activity, the number of words they typed across all questions, and a self-reported rating along a 7-point scale (-3 to +3) of whether they felt the activity was worth their time.
We call these measures \textit{Completion Time}, \textit{Response Length}, and \textit{Perceived Time Commitment} respectively. 
All three measures were compared across conditions using independent samples t-tests.

\subsection{Ethical Concerns}
We addressed potential ethical concerns in the same way we did for Study 1 (Section~\ref{sec:study1ethical}).

\subsection{Results}
\input{figures/study2-table}
The summary statistics for the collected measures are shown in Table~\ref{tab:study2}.

\subsubsection{H1 (Perceived Benefits)}
Participants in the \textit{RQA} condition saw significantly more utility in completing the activity than those in the \textit{Baseline} condition (t(193) = 2.82, p < 0.01).
The average rating for our RQA was 1.2, while the average for the single-question activity was 0.5.
Although both of these averages were near the neutral score of 0, there were many more positive ratings for our RQA.
79\% of the participants who used our RQA gave a non-neutral positive score, while only 57\% did the same for the single-question activity.
Participants also reported a statistically significant change in stress rating in the \textit{RQA} condition relative to the \textit{Baseline} (t(205) = 3.60, p < 0.001).
Whereas people who used our RQA experienced a 0.7-point decrease in their perceived stress rating, people who used the single-question activity actually experienced a 0.4-point increase in theirs.
We conclude from these results that the additional questions from our RQA were not only beneficial in mitigating stress, but also necessary to counteract an initial increase in stress from revisiting the troubling situation.

\subsubsection{H2 (Elapsed Time)}
Participants in the \textit{RQA} condition took 8.9 minutes on average to complete the activity, while those in the \textit{Baseline} condition only took 1.6 minutes on average; the difference between the two conditions according to this measure was statistically significant (t(213) = 7.88, p < 0.001).
We also found that participants wrote significantly longer responses while going through the RQA.
Participants in the \textit{RQA} condition wrote 87 words on average, while those in the \textit{Baseline} condition wrote 29 words on average; that difference was also statistically significant (t(213) = 4.49, p < 0.001).
Although the RQA required significantly more effort, there was no statistically significant difference in people's subjective perception of the activity duration (t(180) = -1.37, p = 0.17).
We conclude from these results that people found value in the additional time they spent completing the series of questions.

%% file: figures/study2-table.tex
\begin{table}[]
\centering
\begin{tabular}{|p{0.3\textwidth}|p{0.15\textwidth}|p{0.15\textwidth}|}\hline

\textbf{Measure} & \textbf{RQA} & \textbf{Baseline}  \\ \hline\hline
Perceived Utility** & 1.2 $\pm$ 0.2 & 0.5 $\pm$ 0.2 \\ \hline
Perceived Stress Change*** & -0.7 $\pm$ 0.2 & 0.4 $\pm$ 0.1 \\ \hline\hline

Completion Time*** & 8.9 $\pm$ 0.8 minutes & 1.6 $\pm$ 0.3 minutes \\ \hline
Response Length*** & 87 $\pm$ 11.6 words & 29 $\pm$ 5.5 words \\ \hline
Perceived Time Commitment & -0.2 $\pm$ 0.2 & -0.3 $\pm$ 0.2 \\ \hline


\end{tabular}\\
\small{* p<.05, ** p<.01, *** p<.001}
\caption{Summary measures and statistics from Study 2. Statistical significance between the measures in \textit{RQA} versus \textit{Baseline} is indicated in the first column according to an independent samples t-test. Averages are given with the standard error within each condition.}
\label{tab:study2}
\end{table}

%% file: chapters/6-longitudinal-study.tex
\section{Study 3: Observing Repeated Engagement with the RQA}
After finding that the RQA delivered short-term benefits when presented to people as a one-time event, we sought to explore how people would perceive the activity during their everyday lives.
We also wanted to explore how to best prompt people to engage with the RQA using two forms of asynchronous and low-cost CMC platforms: email and text messaging. 
Unlike the previous two studies, this one was conducted during the COVID-19 pandemic.

\subsection{Participants}
We recruited 11 participants (8 males, 3 females) with an average age of 20.6 years.
Participants were recruited via email and word-of-mouth from the same university campus community, and there no inclusion criteria.
We refer to these participants as D1--D11.
Participants were not compensated for completing our RQA to avoid undue influence on their level of engagement; however, they were compensated \$10 CAD for completing surveys and \$15 CAD for being interviewed. 

\subsection{Procedure}
Participants were recruited to take part in our study for two weeks.
During the enrollment phase, participants were asked to specify hours during each day when they would prefer to receive a notification to complete the RQA.
They were asked to provide separate preferences for email and text messaging, and they were allowed to select multiple times during a given day.

Participants were then randomized into one of two groups. 
One group received emails on the first week and text messages on the second week, while the other group experienced the reverse. 
The notifications prompted participants to complete the RQA and provided them with a link to the activity.
We used the same link each time, and participants were aware of this fact.
Participants were prompted to complete the activity once per day for up to three days within a given week, which is similar to what has been done in previous work~\cite{beacham2020brief}.
Eight of the participants were available for more than three days, so the days on which they received prompts were randomly selected.
The three participants who were available for fewer than three days (D2, D8, and D9) received a message on every day of their availability.

At the end of the study, participants were asked to complete an exit survey containing questions about their overall experience and their CMC preferences in the context of the RQA. 
They were then invited to a semi-structured interview session to elaborate on their experience.
The interviews lasted 15--30 minutes, with frequent topics including the barriers people faced while completing the RQA, the applicability of the RQA to their lives, and the tradeoffs of being prompted to complete the RQA repeatedly.
Nine people completed the exit survey, and seven people took part in the interviews. 
The interviews were conducted over the Zoom teleconferencing platform.

\subsection{Data Analysis}
We recorded a variety of data to assess how people engaged with our RQA.
We measured how often participants responded to our prompts without a limit on how long they had to respond.
In other words, if a participant received a prompt in the morning but waited until the next day to complete our RQA, we still counted that as a response.
We calculated response rate in this way since it is well-documented that people respond to emails and text messages at their convenience rather than at the moment of reception~\cite{tyler2003can}.
As in Study 2, we asked participants to rate their stress along an 11-point scale before and after the activity, and we report the change in that rating. 
We also report the time it took for participants to complete the RQA and the word count of their responses as proxies for engagement.
We analyzed the interview responses using the same procedure that was applied for Study 1; however, we did so with a new, blank codebook.

\subsection{Ethical Concerns}
We addressed potential ethical concerns in the same way we did for Study 1 (Section~\ref{sec:study1ethical}).

\subsection{Results}
\begin{figure*}
    \begin{subfigure}[b]{.48\columnwidth}
        \includegraphics[width=\textwidth]{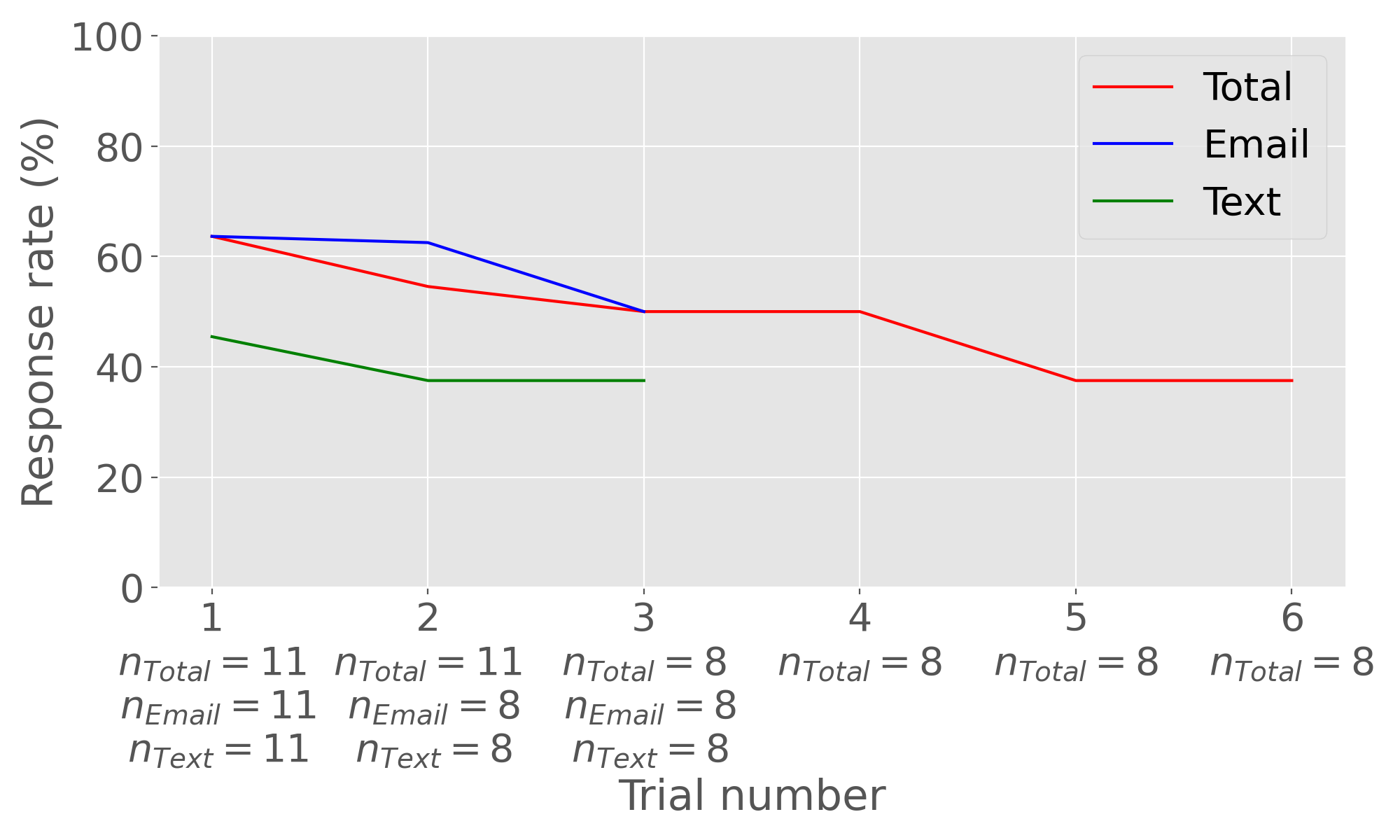}
        \caption{Response rate}
        \label{fig:response_rate}
    \end{subfigure}
    \hfill
    \begin{subfigure}[b]{.48\columnwidth}
        \includegraphics[width=\textwidth]{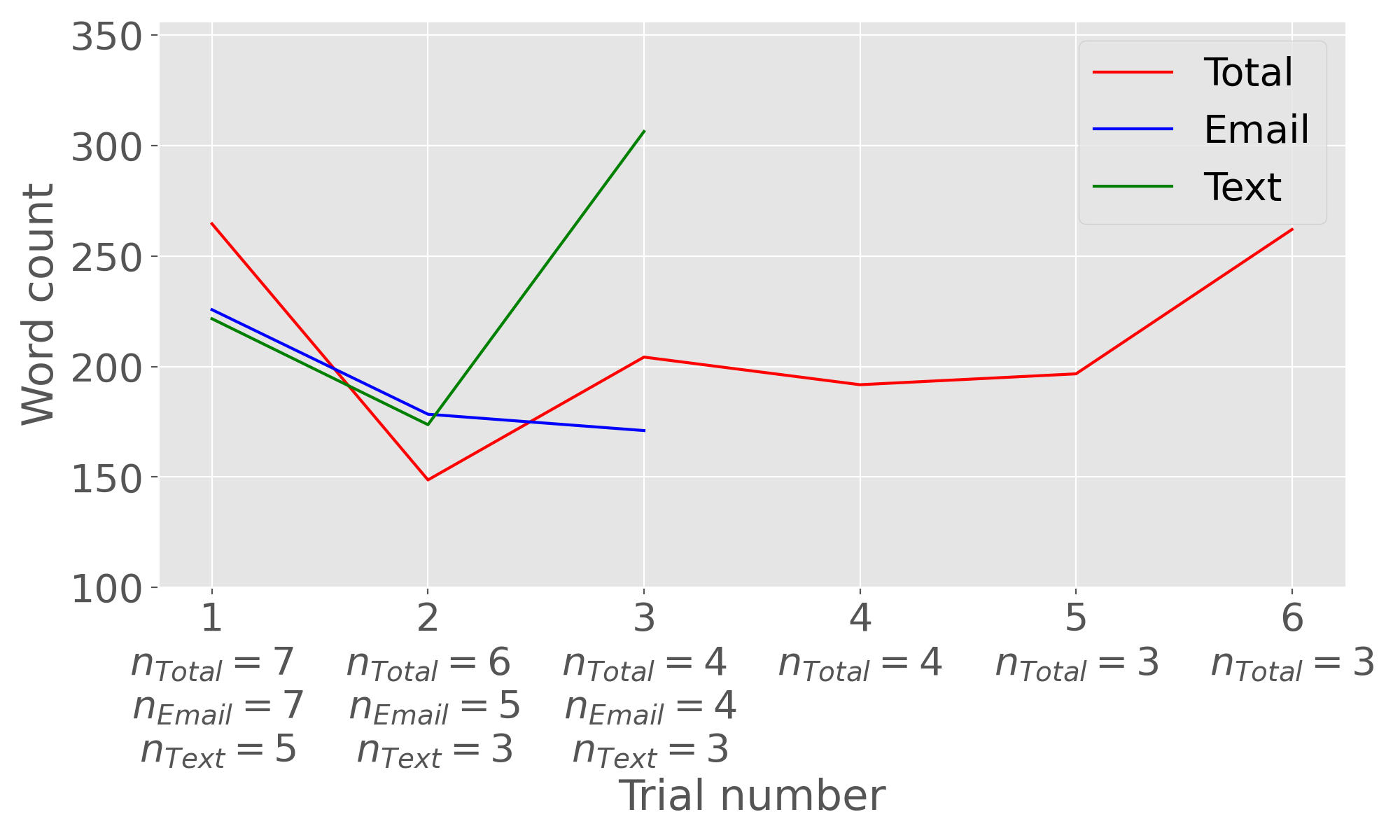}
        \caption{Word count}
        \label{fig:word_count}
    \end{subfigure}

    \begin{subfigure}[b]{.48\columnwidth}
        \includegraphics[width=\textwidth]{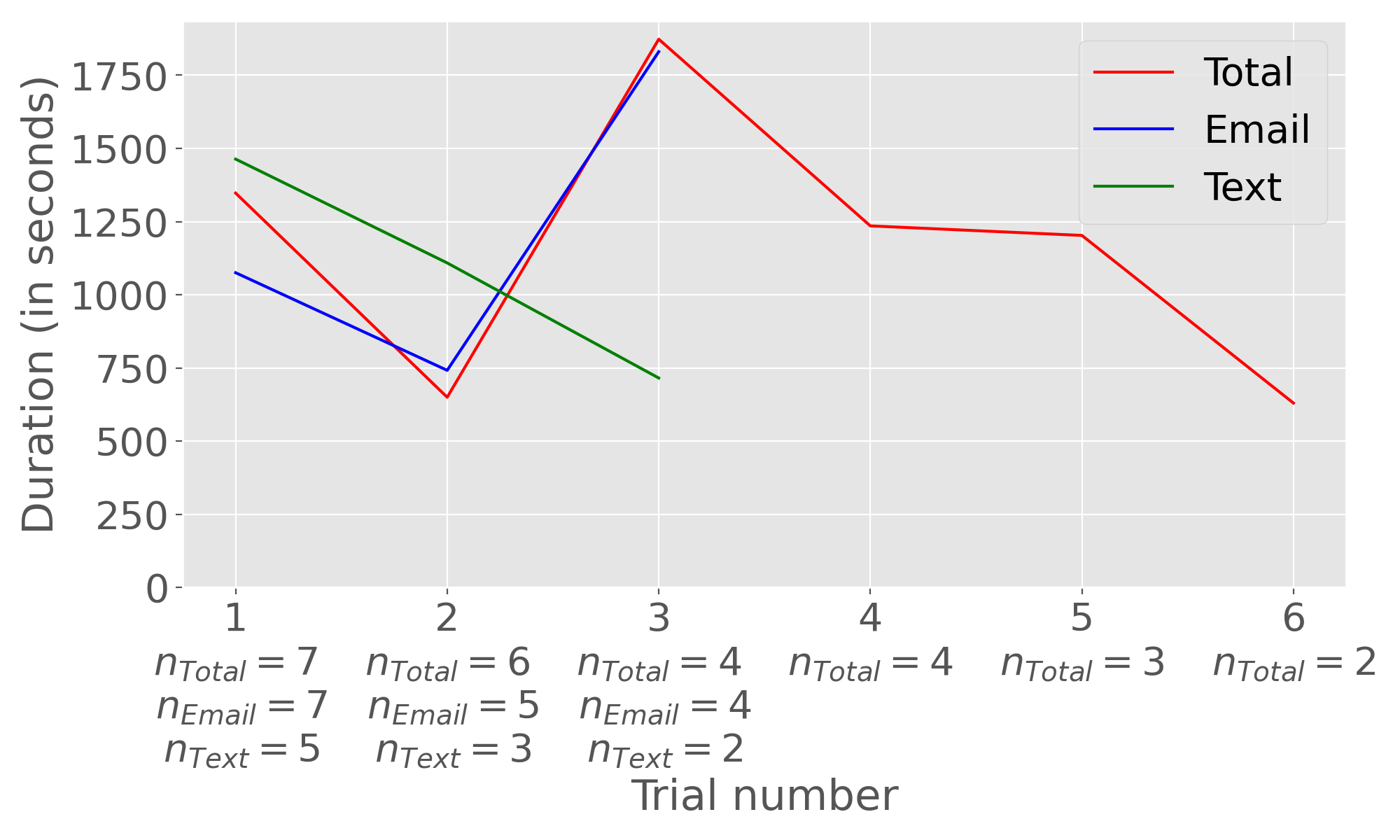}
        \caption{Completion time}
        \label{fig:time_spent}
    \end{subfigure}
    \hfill
    \begin{subfigure}[b]{.48\columnwidth}
        \includegraphics[width=\textwidth]{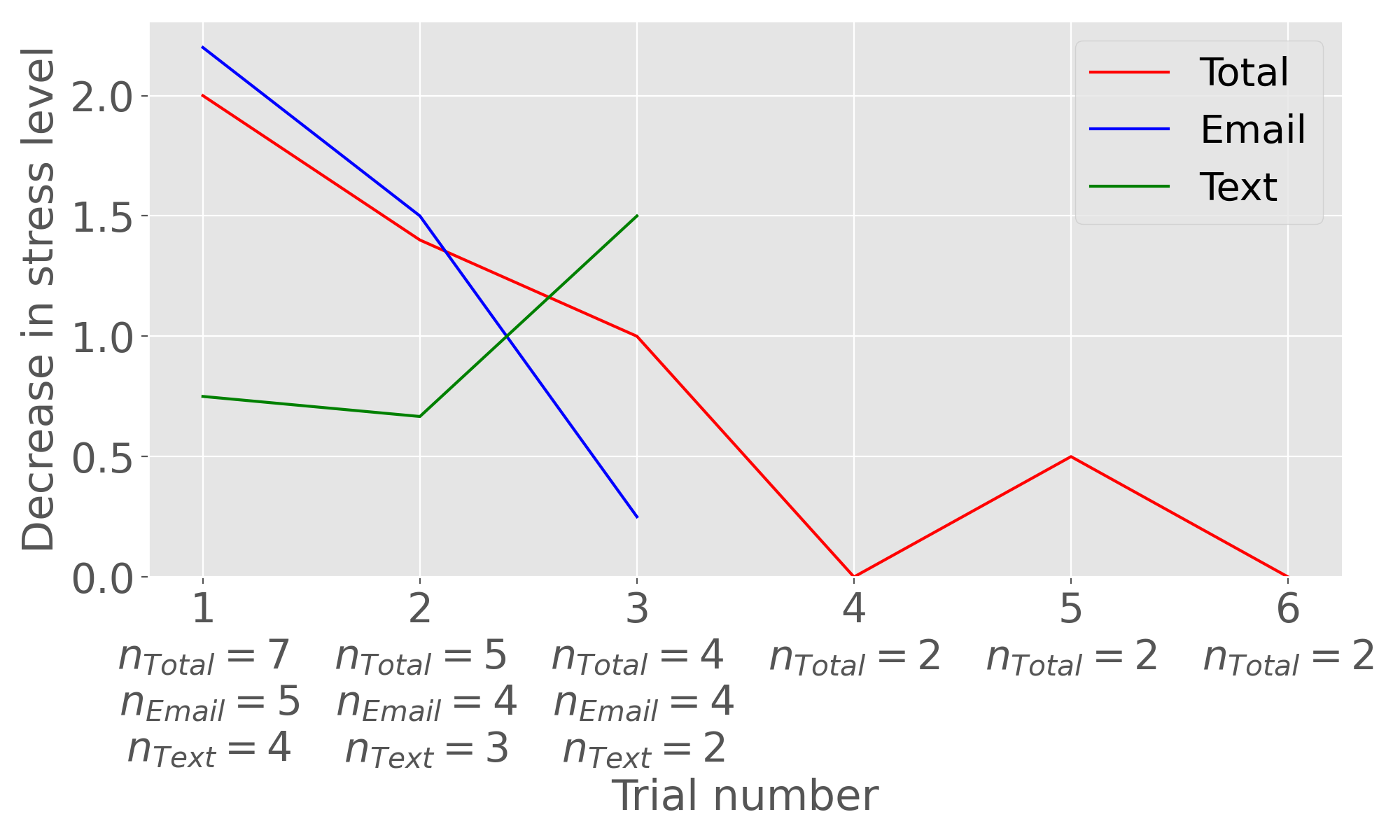}
        \caption{Decrease in self-reported stress level}
        \label{fig:stress_decrease}
    \end{subfigure}
    \caption{Summary plots of the (\ref{fig:response_rate}--\ref{fig:time_spent}) engagement metrics and (\ref{fig:stress_decrease}) decrease in stress level over time. The tick marks below each graph indicate the sample size for each data point, which may vary across measures due to missing data or repeated samples.}
    \label{fig:four_figures}
\end{figure*}

Fig.~\ref{fig:four_figures} summarizes the quantitative data that we collected throughout our study.
The figures not only show the aggregated data across all interactions with our RQA, but also splits the results according to the CMC platform through which the prompts were sent.

\subsubsection{Overall Engagement}
We observed moderate engagement throughout the two-week period of our study. 
We sent participants a total of 54 prompts via email and text messaging (equally distributed between the two), and participants completed the RQA in 27 of those cases (50\%).
On three occasions, participants completed the RQA twice in response to the same prompt, so our RQA was actually completed 30 times during the study.

On average, people spent 18.5 minutes, wrote a total of 212 words, and experienced an average stress level reduction of 1.2 points after completing the RQA. 
Participants were much more engaged with our RQAs in this study compared to the MTurkers in Study 2 (i.e., spent more time and typed longer responses), even as they completed it multiple times. 
One explanation for this discrepancy could be the amount of time participants were willing to commit to the study.
Participants in Study 2 likely completed our RQA in the midst of other crowdsourcing tasks or during their busy workdays.
On the other hand, participants from Study 3 were able to pick a more convenient time later in the day (or a following day), which in turn gave room for a longer time investment.
D3 validated this hypothesis from their experiences:
\italquote{Although I initially said that I would be available in the morning, I found the best time to do it in the time between 9--11 pm. I used to see the emails and text messages short after they came, but I used to only do them at my convenient times in the night.}
Our data also supports this observation, as participants spent an average of $\sim$23 hours between receiving a prompt and completing the RQA.

Nevertheless, our quantitative and qualitative data shows that people could spend as little or as much time as they wanted with the activity without the need for significant scaffolding.
In the interviews, participants expressed similar opinions about the benefits of our RQA as they did in our previous studies. 
Most notably, participants saw benefits to having a structured way of organizing thoughts as it helped them identify triggers and devise an alternate way of thinking.

\subsubsection{Repeated Engagement with the RQA}
One major goal of this study was to observe how participants engaged with the RQA over time. 
Unsurprisingly, we observed that the response rate decreased over time. 
Fig.~\ref{fig:response_rate} shows that the response rate was 63.6\% when participants received their first prompt and then 54.5\% for the second prompt. 
By the time they had seen six prompts, the response rate went all the way down to 37.5\%.
When we asked participants to explain this trend during our interviews, the main complaint was that doing the same activity in such a short interval was boring and tedious. 
D3 mentioned that the length of the activity was acceptable for a one-time event, but when they had to do the activity thrice in the same week, it ``came across as a chore''.
Similarly, D10 said,
\italquote{When it started coming every other day, I felt like I had to do a school homework. So I felt a little bit of pressure to do the activity.}
When we asked participants how often they would want to go through the RQA, only one person wanted see a prompt every other day.
Four people said that they would want to get prompts once a week, one person said once every two weeks, and three people said once a month.
The preference towards longer intervals was not only due to the monotony of the task, but also because participants struggled to think of new troubling situations to reflect upon. 
D6 remarked,
\italquote{By the time I got the last prompt, I could not find a stressful situation in my life. Maybe the frequency should vary depending on the amount of stress a person is going through.}

Despite these concerns, participants acknowledged that prompting them to do our activity on a regular basis had value. 
A few participants noted that they would have forgotten to revisit the RQA had they not been given periodic reminders.
D1 also believed that they ``got more comfortable with the activity [over time] and started setting aside a time to do the activity''.

Although the response rate decreased over time, Fig.~\ref{fig:word_count} shows that those who stuck with our RQA wrote similar amounts of text over time.
The amount of time that people spent with the activity was highly variable throughout the study, which can largely be attributed to the low sample size, but there were no clear monotonic trends in the data.
However, Fig.~\ref{fig:stress_decrease} shows that people experienced smaller improvements in stress level as they repeatedly went through the RQA. 
The most common explanation that participants gave during our interviews was that whenever they reflected on the same troubling situation, they became less worried about it each time and thus there was less room for improvement.
Fortunately, we did not see any instances when the RQA increased participants' stress level.

\subsubsection{CMC Platforms}
Another goal of this study was to gain insights about the role of technology in deploying RQAs. 
Fig.~\ref{fig:response_rate} shows that there was a noticeable difference between the response rates over email and text messaging. 
Even in our exit survey, eight participants said that they would prefer email over text messaging for doing this activity, and the rest of the participant had no preference. 
One of the main reasons for the preference was the affordances of desktop and laptop computers when it came to completing the RQA.
Most notably, participants commented on how computers are better suited for reading and writing longer passages of text.
D1 said,
\italquote{Typing is very difficult in mobile phones. The screen size is small and editing stuff is a nightmare. On the other hand, if you want to write a long answer, you would probably do that on computer because the process is just much easier.}
Participants also felt that doing the RQA on a computer minimizes the chance for distractions. 
For example, D7 commented that sitting in front of their computer gave them the ``right mindset to do the activity''. 
With a computer, they felt that they had control over their workspace since they could easily close other tabs and applications.
When they tried to do the activity on their phone, on the other hand, there were cases when a call or push notification disrupted their train of thought.

Although email was generally preferred for completing the RQA, many people agreed that mobile phones are a great mechanism for sending notifications and reminders. 
Participants like D4 and D6 had the concern that people may not check their emails as frequently as they check text messages, with D4 saying, 
\italquote{Most of the times, I have my phone in my hand, whereas I check my emails at most once or twice a day. So if you need me to do something immediately, you would probably need me to reach via text messages. I can respond to an email even two days later.}
Participants also informed us of instances when they switched between the two CMC platforms.
When D6 was prompted to do the RQA over text messaging, they sent the link to themselves over social media and then accessed it on their desktop to complete the RQA. 
Some participants suggested that the RQA could actually be advertised over social media platforms like Facebook or Instagram since people normally access their accounts across multiple devices.
In doing so, people could have the option to choose whichever platform they see fit. 
Alternatively, some participants thought that the two CMC platforms could be integrated into the same system.
D6 had the following suggestion:
\italquote{What you can do is you can ask me to answer the questions in the text message, but at the same time you will also send me an email that has the links to the actual page.}



%% file: chapters/7-discussion.tex
\section{Discussion}
In this work, our aim was to create a brief digital intervention that people could complete on their phone or computer to lessen their concerns about a troubling situation.
Our second study demonstrated that the RQA delivered significant benefits relative to a baseline with a commensurate time commitment. 
Meanwhile, our first and third studies elicited qualitative findings that we hope will inform the design of future interventions in this space. 
Most notably, we found that participants appreciated our RQA for its ability to help them to undergo a structured analysis of their troubling situation, identify solutions to improve their situation, and vent out their negative feelings. 
Although we observed that participants felt that the series of questions was worth the additional time commitment, we also saw numerous obstacles towards the long-term engagement with the RQA.
In the following subsections, we elaborate on the implications of our findings, the limitations of our work, and potential directions for future research.

\subsection{Improving the Design of Our RQA}
The success of our RQA does not mean that it could not be improved. 
Although we never observed an increase in participants' stress level after completing the RQA, some participants from Study 1 remarked that the activity left them confused and without a concrete solution.
We hypothesize that such concerns could be remedied by providing users with sample responses to each question as a source of inspiration.
These examples could be curated by researchers, or they could be collected from previous users who voluntarily contribute their responses to a database.
Topic modeling~\cite{blei2003latent} could be used to tag the examples with keywords on subject matter, and an information retrieval system could rank the relevance of these examples. 
Collaborative filtering~\cite{schafer2007collaborative} could even be used to gradually collect ratings for each example and then cater examples to individual users' preferences.

Another way that we could improve our RQA is by personalizing the questions themselves.
The activity could ask users to rate the perceived benefit of each question, or we could automatically assess the quality of each question by examining the corresponding response length.
Using this information, we could extend or emphasize questions that individuals find most beneficial. 
We could also use this information to remove questions that induce stress; however, thought records and behavioral chaining are intentionally designed processes with many critical steps, so removing questions may detract from the activity's benefits.

\subsection{Alternative RQAs}
Our nine-question RQA that took inspiration from CBT, particularly thought records and behavioral chaining.
However, future work could investigate RQA designs or brief CMC-mediated interventions more broadly based on other psychological principles. 
For example, encouraging expressions of gratitude or social connections with others can play a key role in stress and depression management~\cite{brown2003providing, cheng2015improving}, so RQAs built around those practices can similarly help people manage their well-being.
Future work could also explore different activity structures.
Some participants in Study 3 complained about the inconvenience of typing on their smartphones, so an alternative activity could ask people to record and listen to their voice for reflection.
Another activity could encourage peer support by starting conversations between online peers.
Lastly, HCI researchers could create brief activities centered around other psychological frameworks beyond CBT, with past examples being centered around mindfulness~\cite{quinones2019reducing}, motivational interviews~\cite{dunn2001use}, and acceptance and commitment therapy (ACT)~\cite{fabricant2013comparison}. 

\subsection{Considerations for Long-Term Engagement}
Our two-week deployment in Study 3 enabled us to gain insights on how people would engage with the RQA over a period of time.
Although participants were pleased with the fact that they could specify their hours of availability, receiving prompts for the RQA three times within the same week was overwhelming for most people.
The biggest criticism was that people received multiple prompts without experiencing a new troubling event, so they either had to go through the RQA while analyzing the same event as before or recalling a troubling event from the distant past.
Ideally, the frequency of prompts would adapt dynamically according to a person's needs.
One participant suggested that users should have control over how often they receive reminders to complete the RQA, explaining that individuals who experience more stress than others might benefit more from doing these activities in short intervals.
Going a step further, future work could integrate physical activity trackers, smartphone sensors, and IoT devices to automatically detect periods of heightened stress~\cite{padmaja2018machine,ng2018veterans}, turning the RQA into a just-in-time adaptive intervention. 

Another issue with completing our RQA too often was that answering the same set of questions became boring and tedious, yet adjusting the prompt frequency alone may not be enough to resolve these concerns.
One way to add variety would be to mix our RQA with other microinterventions, as was done by \citet{paredes2014poptherapy} in their PopTherapy work.
Brief interventions like our RQA could also serve as a gateway to more time-consuming exercises or professional therapy.
By giving people a preview of the potential improvement in mood they can receive from articulating their thoughts and emotions, habits can be formed and users may become more motivated to build on that momentum~\cite{eyal_2019}.

\subsection{Limitations}
Rather than developing a mental health intervention for people suffering from clinical depression or other psychological disorders, our intention was to design our RQA for as broad of a population as possible.
We did this by asking participants to think through a troubling situation --- something that everyone experiences at some point in their life. 
However, it would be imperative for researchers to conduct further studies specifically with more vulnerable populations to understand the benefits and potential risks of digitally delivered RQAs.
We suspect that self-reflection could serve as a convenient mechanism for people to practice what they learn in psychotherapy, but it could also perpetuate negative thought patterns. 
We also recognize that our participant cohorts --- MTurk crowdworkers and university students --- do not represent all aspects of the general public. 
Most of our qualitative findings were not tied to participants’ specific contexts, and we did not find any obvious evidence of substantial differences between the cohorts. 
Nevertheless, future work could deploy RQAs to more diverse cohorts. 

%% file: chapters/8-conclusion.tex
\section{Conclusion}
In this work, we used principles from CBT to design questions that help people articulate, reflect on, and change their thoughts and emotions about a troubling situation.
The three studies we presented in our paper provide evidence that people are willing to engage with and find value in brief self-reflection activities delivered through CMC platforms, even without scaffolding like training or real-time feedback. 
We found that our RQA not only helped people reduce their stress, but also helped them challenge their potentially negative thought patterns and identify alternative ways of thinking.
We also found that people were willing to use the RQA more than once, although future work is needed to strike a balance between utility and monotony.
We hope that our work inspires other researchers to explore new formats for brief interventions that help people with their everyday struggles.